\documentclass{jpconf}

\usepackage{amssymb}
\usepackage{graphicx}        % standard LaTeX graphics tool
\usepackage{slashed} % EXTRA PACKAGE --- Emelyanov

\makeindex             % used for the subject index
\begin{document}
\title{QED loop effects in the spacetime background  of a Schwarzschild black hole}
\author{Viacheslav A. Emelyanov}
%\institute{V.A. Emelyanov \at  Institute of Theoretical Physics, Karlsruhe Institute of Technology, \\ 76131 Karlsruhe, Germany \\ \email{viacheslav.emelyanov@kit.edu}}
%\author{V.A. Emelyanov}
\address{Institute of Theoretical Physics, Karlsruhe Institute of Technology,  76131 Karlsruhe, Germany}
\ead{viacheslav.emelyanov@kit.edu}
%\maketitle

%\vspace{-15mm}

\begin{abstract}
The black-hole evaporation implies that the quantum-field propagators in a local Minkowski frame acquire a correction,
which gives rise to this process. The modification of the propagators causes, in turn,
non-trivial local effects due to the radiative/loop diagrams in non-linear QFTs. In particular, there
should be imprints of the evaporation in QED, if one goes beyond the tree-level approximation.
Of special interest in this respect is the region near the black-hole horizon, which, already at tree level, appears
to show highly non-classical features, e.g., negative energy density and energy flux into the black hole.
\end{abstract}

%\vspace{-5mm}

\section{Introduction}
\label{sec:Introduction}
After many years of theoretical and experimental studies, we have been successful in establishing a couple of
models, which incredibly well describe physics at microscopic and macroscopic scales. The basic ideas behind
of the Standard Model of Particle Physics and Cosmology, are different, but not
inconsistent. It is the Einstein equivalence principle that allows us to go over from the general relativity regime
to the particle physics one.

According to the large scale observations, the global geometrical structure of the Universe
turns out to be roughly described by de Sitter spacetime. At smaller length scales,
the Universe is no longer homogeneous and isotropic as a consequence of the presence of Dark Matter as
well as baryonic matter in the form of clusters of galaxies, galaxies and so on. However, Minkowski spacetime
appears to be a good approximation to the Universe locally, i.e. in the neighbourhood 
of any given space-time point.

The geometrical structure relevant for Particle Physics is given by the Minkowski metric of a local
Minkowski frame. The Poincar\'{e} group
$\textrm{R}^4{\rtimes}\textrm{SO}(1,3)$ is, thus, a local isometry group of the Universe. This group is employed to
define the notion of particles, which are associated with its unitary and irreducible representations,
with a no-particle state being the local Minkowski vacuum. The
Lehmann-Symanzik-Zimmermann reduction formula relates, in turn, physical particle states with poles of
the quantum-field propagators. We know \emph{a posteriori} that this formalism adequately describes various
reactions observed in the particle colliders on Earth.

The notion of particles identified with the unitary, irreducible representations of the Poincar\'{e} group 
and, hence, labeled by values of its two Casimir generators,
i.e. mass and spin, is the only one, which has been successfully tested so far. The Wigner particle is, thus,
a localized object, that can be defined in a local Minkowski frame only. It seems that if the local curvature
in a certain space-time region of the Universe becomes comparable with, e.g., the Compton length $\lambda_c$
of the electron, then one should not expect that the notion of the electron is physically meaningful there.
Nevertheless, the Wigner electrons (i.e. electrons we observe in the particle colliders)
might, still, make physical sense if relativistic, i.e. correspond to the de Broglie length $\lambda_e \ll \lambda_c$.
Therefore, the Wigner electrons have to be highly relativistic in that region.
Although the proof of this conjecture in a lab is impossible, we take it for granted below as a natural
extrapolation of physics we have experimentally probed up to now.

The fact that we find ourselves in non-flat spacetime implies that one should bear in mind
the curvature length scale $l_c$. Still, the Minkowski-space approximation employed
in Particle Physics is adequate, whenever the length scale of a certain particle reaction is negligible with respect
to $l_c$. Consequently, one does not need to integrate over all space-time points of the Universe
in the coordinate representation of vertices in non-linear QFTs. Moreover, the same also appears to hold
in the path integral formalism and, furthermore, one can use the powerful technique of the Fourier transform
as if the Universe were globally Minkowski spacetime.

Thus, one can imagine that the local curvature at a given point of the Universe is described by the length scale
$l_c$. According to the Einstein equivalence principle, one can introduce a local Minkowski frame in its vicinity. In this
frame, we can follow standard methods used in Particle Physics to quantize, e.g., the Maxwell or Dirac field. There
is no reason to expect that this procedure
is physically meaningless, because otherwise it would be unnatural to assume that the formalism works on Earth only.
Of course, one should pay attention to the fact that the Standard Model is not scale invariant. For instance,
if $l_c \lesssim \lambda_c \propto 1/m_e$, one should then allow only electrons/positrons as
external legs in the Feynman diagrams with momenta $|\vec{p}| \gg 1/l_c$.
In this paper, we intend to employ these ideas near a black hole, so that $l_c$ corresponds to
the size of the black-hole event horizon, i.e. $l_c = r_H = 2M$, where $M$ is a black-hole mass. 

Throughout this paper we assume that $c=G=k_B = \hbar = 1$, unless otherwise stated.

%\vspace{-5mm}

\section{Loop effects in Quantum Electrodynamics}
\label{sec:2}
\begin{figure}[t]
%\sidecaption
% For example, with the graphicx style use
    \centering
\includegraphics[scale=.51]{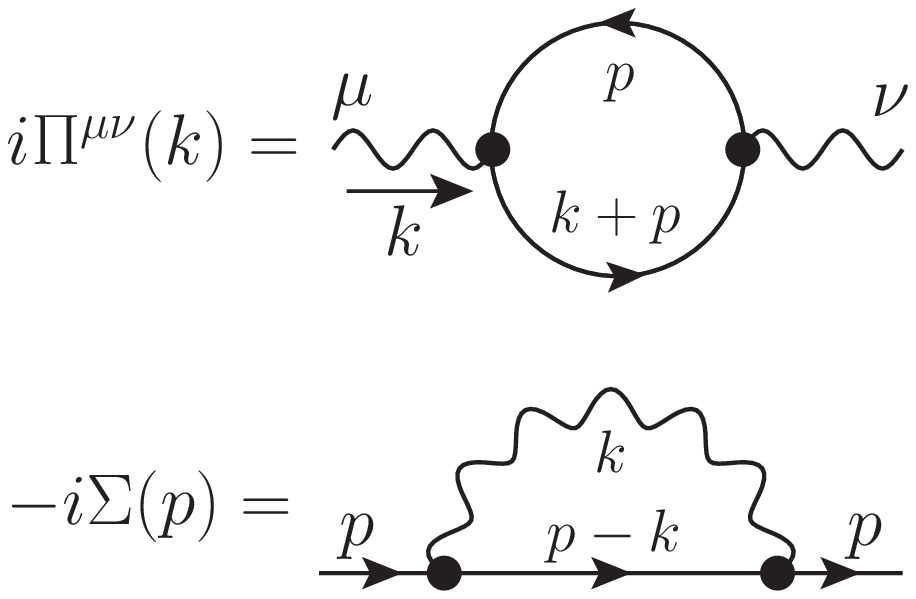}
%
% If no graphics program available, insert a blank space i.e. use
%\picplace{5cm}{2cm} % Give the correct figure height and width in cm
%
\caption{One-loop diagrams providing the lowest order correction to the photon as well as electron self-energy.
The object $\Pi^{\mu\nu}(k)$ is known as the vacuum polarization tensor. The vacuum polarization gives rise to
the running of the electric charge and contributes to the Lamb shift.}
\label{fig:1}       % Give a unique label
\end{figure}

A plenty of observable effects in QFT cannot be understood at tree level, i.e.
without counting loop diagrams. In quantum electrodynamics, the most remarkable manifestations 
of the loops are the anomalous magnetic moment of the electron, the running of the electric charge and
the Lamb shift. We shall deal in this paper with the one-loop correction to the photon and electron
self-energy (see Fig.~\ref{fig:1}).

%\vspace{-5mm}

\subsection{Isotropic electron-positron plasma}
\label{subsec:2}
In the presence of particles and/or external fields, novel non-trivial manifestations of the loops appear.
As an instructive example, we consider first an isotropic and neutral plasma composed of electrons, positrons and
photons.

%\runinhead{Maxwell- and Dirac-field propagators}
\paragraph{Maxwell- and Dirac-field propagators:}
To determine how the plasma particles influence, e.g, the photon kinematics, one
needs to know the vector- and fermion-field propagators inside the plasma. One can show that
\begin{eqnarray}\label{eq:1}
\underbrace{\frac{i(\slashed{p} + m)}{p^2-m^2+i\varepsilon}}_\textrm{\small Vacuum} & \;\;\Longrightarrow\;\; &
\underbrace{\frac{i(\slashed{p} + m)}{p^2-m^2+i\varepsilon} -
2\pi\,\frac{(\slashed{p}+m)}{e^{\beta|p_0|}+1}\,\delta\big(p^2-m^2\big)}_\textrm{\small Isotropic plasma}\,,
\end{eqnarray}
where $\beta \equiv 1/T$ is the inverse plasma temperature, and
\begin{eqnarray}\label{eq:2}
\underbrace{\frac{-i\eta_{\mu\nu}}{k^2+i\varepsilon}}_\textrm{\small Vacuum} & \;\;\Longrightarrow\;\; &
\underbrace{\frac{-i\eta_{\mu\nu}}{k^2+i\varepsilon} -
2\pi\,\frac{\eta_{\mu\nu}}{e^{\beta|k_0|}-1}\,\delta\big(k^2\big)}_\textrm{\small Isotropic plasma}\,.
\end{eqnarray}

Since the electron-positron field is massive, the extra term in its propagator is not suppressed
by the Boltzmann factor $\exp(-\beta m_e)$ if we consider the case $T \gg m_e$.
It is legitimate, then, to neglect the mass parameter $m_e$, so that the fermion field becomes
effectively massless. This limit is known as the hard-thermal-loop approximation~\cite{LeBellac}.

%\runinhead{Modified dispersion relation}
\paragraph{Modified dispersion relation:}

The one-loop effects originating from non-vanishing temperature are the Debye screening of
a point-like charge, emergent propagating degrees of freedom (so-called plasmon and plasmino) as a result of
the collective plasma-particles excitations and others (see, e.g.,~\cite{LeBellac}). We focus herein on the
photon and electron dispersion relations.

With the propagators given in Eqs. (1) and (2), we find in the on-mass-shell limit in the hot plasma ($T \gg m_e$) that
\begin{eqnarray}\label{eq:3}
k_0^2 - \vec{k}^2 &\approx& \frac{2\pi}{3}\,\alpha T^2\,,
\\[-1.0mm]\label{eq:4}
p_0^2 - \vec{p}^2 &\approx& m_e^2 + \pi \alpha T^2
\end{eqnarray}
at one-loop level, where $\alpha \equiv e^2/4\pi$ is the fine-structure constant and $m_e$ is a physical electron
mass (whereas $m$ is a bare electron mass). Thus, QED photons acquire a thermal gauge-invariant mass, which
changes their kinematics~\cite{LeBellac}.

%\vspace{-5mm}

\subsection{Anisotropic electron-positron plasma}
\label{subsec:3}
An anisotropic plasma is characterized by a unit vector $\vec{n}$, that gives the direction
along which the plasma moves. This medium appears to be closely related to the quantum physics near
to and far from the black-hole horizon~\cite{Emelyanov-1,Emelyanov-2,Emelyanov-4,Emelyanov-5}.

%\runinhead{Maxwell- and Dirac-field propagators}
\paragraph{Maxwell- and Dirac-field propagators:}

If the anisotropic and neutral electron-positron plasma is held at temperature $T$, then
one needs to implement in Eqs. (1) and (2) the following changes:
\begin{eqnarray}\label{eq:5}
\delta\big(k^2\big) & \;\;\Longrightarrow\;\; & 2\pi |\vec{k}|\,\delta\big(\vec{k} - k_0\vec{n}\big)\,,
\\[-1.0mm]\label{eq:6}
\delta\big(p^2 - m^2\big) & \;\;\Longrightarrow\;\; & 2\pi |\vec{p}|\,
\delta\big(\vec{p} - (p_0^2-m^2)^\frac{1}{2}\vec{n}\big)\,,
\end{eqnarray}
in order to gain the photon- and electron-positron-field propagators in this medium.

Starting with quantum kinetic theory (see, e.g.,~\cite{Emelyanov-3}), it is straightforward to show that the anisotropic plasma is
described by the following one-particle distribution:
\begin{eqnarray}\label{eq:7}
f(x,p) &=& \frac{g_s}{2\pi^2}\frac{|\vec{p}|^2}{e^{\beta p_0} - (-1)^{2s}}\,\theta\big(\vec{p}{\cdot}\vec{n}\big)
\delta\big(\vec{p}{\times}\vec{n}\big)\,,
\end{eqnarray}
where $p_0$ is either $|\vec{p}|$ or $(|\vec{p}|^2 + m_s^2)^\frac{1}{2}$, depending on whether the quantum field of spin
$s$ is massless or massive and $g_s$ is its number of degrees of freedom. For instance, this implies that the photon number current
$\vec{j}_\gamma$ is given by $n_\gamma{\cdot}\vec{n}$, where $n_\gamma = 2\zeta(3)T^3/\pi^2$ is the photon
number density in the plasma. Therefore, the anisotropy of the plasma manifests itself in the macroscopic local
observables.
 
%\runinhead{Modified dispersion relation}
\paragraph{Modified dispersion relation:}

It turns out that the modified dispersion relations given in Eqs. (3) and (4) are independent on
whether the electron-positron plasma is isotropic or anisotropic. This holds only at one-loop
approximation. The photon dispersion relation at two-loop order gets a dependence on the unit vector
$\vec{n}$, implying that the two-loop diagrams do not contribute to the thermal photon
mass~\cite{Emelyanov-6}.

%\vspace{-5mm}

\subsection{Near-horizon region of small black holes}
\label{subsec:4}

The near-horizon region of black holes is exotic from a classical point of view. Specifically, the renormalized stress
tensor of the quantum
fields describes an inward flux of the negative energy density, originating well outside of the
black-hole horizon~\cite{Unruh,Candelas,Giddings-3}. This explains, however, why the
black-hole horizon shrinks, i.e. $2\dot{M} < 0$, where dot refers to the Schwarzschild time coordinate.

This non-classical flux, like the Hawking flux in the far-from-horizon region, is anisotropic. This phenomenon
is described
by the field propagators of the form given in Eqs. (1) and (2) with the modifications (5) and (6) implemented
with an extra multiple factor roughly equaling to $-27/4$ and
$\beta = 2/T_H$~\cite{Emelyanov-3,Emelyanov-4},\footnote{The factor $2$ in $\beta$
comes from the Fermi normal coordinates, which are introduced in the near-horizon region.}
where $T_H$ is the Hawking temperature. Thus, the following result holds:
\begin{eqnarray}\label{eq:8}
k_0^2 - \vec{k}^2 &\approx& -\frac{9\pi}{8}\,\alpha T_H^2\,,
\\[-1.0mm]\label{eq:9}
p_0^2 - \vec{p}^2 &\approx& m_e^2 -\frac{27\pi}{16}\,\alpha T_H^2\,,
\end{eqnarray}
for black holes with $T_H \gg m_e$~\cite{Emelyanov-5}.

\section{Concluding remarks}
\label{sec:3}

We have found above that $k^2 < 0$ and $p^2 < 0$ close to the event horizon. What is the physics of this result?

%\runinhead{Einstein acausality vs. instability}
\paragraph{Einstein acausality vs. instability:}

First, it is useful to consider a classical field model with a ``wrong" sign in the mass-squared term.
In general, there are two logically non-excludable ways of quantizing this theory. One can
either discard modes with imaginary frequencies or take those into
account. In the former case, the Einstein causality is violated, whereas, in the latter one, it is not, but
the price is the quantum-vacuum instability~\cite{Dhar&Sudarshan,Schroer, Aharonov&Komar&Susskind}.
The very reason of the instability are frequency modes with $|\vec{q}| < \mu$, where $\mu > 0$
is the field ``mass". In other words, the modes with $q_0^2 = \vec{q}^2 - \mu^2 < 0$ are
exponentially increasing in time. This model, however, is not equivalent to our case, because
$m_\gamma^2 < 0$ is a one-loop result obtained within standard QED, but in the non-standard
quantum-vacuum state.

It is further instructive to consider a classical field model with the standard mass term, but, e.g., in flat de Sitter
spacetime with the Hubble constant $H$. In this case, the superhorizon modes increase without
bound at future time infinity. These correspond to imaginary time-dependent frequencies. This fact does
not serve any problem in Particle Physics, because those modes have a physical wavelength larger than the
space-time curvature, i.e.
no notion of particles can be associated with them~\cite{Mukhanov&Winitzki}.

It, now, seems that the tachyonic scalar field model with $\mu \lesssim H$ could be physically (almost)
indistinguishable from a massless scalar field in local tests in the particle colliders. Still, the dispersion 
relation is $q_0^2 = \vec{q}^2 - \mu^2$ with the IR cutoff $l_{IR} \ll 1/H \lesssim 1/\mu$
and, hence, $|\vec{q}| \gg H \gtrsim \mu$. This, in turn, means that
$dq_0/d|\vec{q}| \approx 1 + \mu^2/2\vec{q}^2 > 1$ holds for scalar particles. Moreover, any local perturbation
(e.g., a particle) is oblivious to the large-scale structure and, thus, described by momenta with $|\vec{q}|$
much larger than
$1/l_{IR}$,\footnote{It should be noted that the (standard) Fourier transform is inapplicable
at length scales, at which the space-time curvature is non-negligible. This is, however, not a problem
in Particle Physics, because particles are localized in coordinate space. The infinite-extent
approximation of a local Minkowski frame is, thus, applicable, if particles are characterized
by a high-enough energy scale.}
implying that this perturbation is described by the group velocity larger than $c$.

This last example reflects our case, if one takes into account that $m_\gamma^2 < 0$ emerges 
due to the black-hole evaporation at one-loop approximation. Hence, it does not hold at
length scales larger than $r_H$, because the field propagators, we have employed in the one-loop
computations, are accurate in the local Minkowski frame only. Thus, a high-energy QED photon/electron
might be able to propagate out of the black hole, bringing information about its internal structure. This appears
to provide a semi-classical mechanism for the proposal of~\cite{Giddings-1,Giddings-2} towards the
resolution of the information-loss problem.

%\runinhead{Black-hole explosions}
\paragraph{Black-hole explosions:}

The non-violent Einstein acausality near the event horizon implies that the black-hole explosions happen
much earlier, i.e. rather before black holes reach the Planckian size. It might provide the explanation of why
these explosions have not been observed so far, because these appear to be less energetic events in the
sky than it has been thought.

%\begin{acknowledgement}
\ack
It is a pleasure to thank the participants of the 3rd Karl Schwarzschild Meeting
on Gravitational Physics and the Gauge/Gravity Correspondence
for interesting discussions and the organizers for making it all happen.
I am also thankful to Frans Klinkhamer and Jos\'{e} Queiruga for their comments on an early version of this paper.
%\end{acknowledgement}
%
%\vspace{-8mm}

\section*{References}
%%%%%%%%%%%%%%%%%%%%%%%% referenc.tex %%%%%%%%%%%%%%%%%%%%%%%%%%%%%%
% sample references
% %
% Use this file as a template for your own input.
%
%%%%%%%%%%%%%%%%%%%%%%%% Springer-Verlag %%%%%%%%%%%%%%%%%%%%%%%%%%
%
% BibTeX users please use
% \bibliographystyle{}
% \bibliography{}
%

\end{document}